  \documentclass[prl,aps,twocolumn,showpacs]{revtex4}
\usepackage[dvips]{graphicx}
\usepackage{dcolumn}
\newcommand{\be}{\begin{equation}}
\newcommand{\ee}{\end{equation}}
\newcommand{\bea}{\begin{eqnarray}} \newcommand{\ba}{\begin{array}}
\newcommand{\eea}{\end{eqnarray}}    \newcommand{\ea}{\end{array}}

\begin{document}

\title{Negative differential resistance in molecular junctions: The effect
of the electrodes electronic structure} 

\author{ Natalya A. Zimbovskaya$^{1,2,3}$ and Mark R. Pederson$^3$}

\affiliation{$^1$Department of Physics and 
Electronics, University of Puerto Rico-Humacao, CUH Station, Humacao, PR 00791,}
\affiliation{$^2$Institute for Functional Nanomaterials, University of Puerto 
Rico, San Juan, PR 00931,}
\affiliation{$^3$code 6390, Naval Research Laboratory, 4555 Overlook Ave SW, 
Washington, DC 20375}

\begin{abstract}
We have carried out calculations of electron transport through a 
metal-molecule-metal junction with metal nanoclusters taking the part of
electrodes. We show that negative differential resistance peaks
could appear in the current-voltage curves. The peaks arise due to narrow features
in the electron density of states of the metal clusters. The proposed analysis
is based on the ab initio computations of the relevant wave functions and
energies within the framework of the density functional theory using NRLMOL
software package.
\end{abstract}

\maketitle
\date{\today}

Electron transport through molecular-scale systems has been intensively studied
 in the past two decades \cite{1}. Largely, 
the unceasing efforts of the research community to further advance these studies 
are due to important application potentials of single molecules as active 
elements in various nanodevices intended to complement current silicon based 
electronics \cite{2}. 
Among various  important properties of the electron transport 
through metal-molecule junctions one may separate out the negative 
differential resistance (NDR), that is the decrease of the current $I$ while 
the bias voltage $ V$ across the molecule increases. The NDR efffect 
was originally observed in tunneling semiconducting diodes \cite{3,4}. 
Later, the NDR was viewed in quantum dots\cite{5,6} and metal-molecule-metal 
junctions  (see e.g. Refs. \cite{6,7,8,9,10,11,12,13,14}).

Several possible scenarios are proposed to explain the NDR occurence in 
the electron transport through molecules. The NDR could appear due 
to alignment and subsequent misalignment of the Fermi levels of the leads 
with molecular orbitals which happens as the bias voltage increases 
\cite{11,15}. This
may noticeably modify the coupling of the molecule to the leads. The 
variations in the coupling strengths could serve as an immediate
reason for the NDR peaks to appear.
Also, such peaks could occur
as a Coulomb blockade induced effect \cite{5,16}, and/or they could  originate 
from conformational changes in the molecule \cite{17} and electron interaction
with the molecule vibrational modes \cite{7,18}. It is likely that different
mechanisms could play a major part in the NDR appearance in different molecular
junctions where it was observed so far. However, reviewing the available 
experimental data, one may conclude that the most distinguished NDR features
in the current-voltage curves (sharp and narrow peaks separated by intervals of
extremely low conductivity, like those reported in the Ref. \cite{8}) are usually
attributed to the matching-mismatching of the molecule energy levels with 
those of the leads. Keeping in mind that precisely
such NDR characteristics are potentially valuable for molecular electronics 
applications, we further concentrate on this mechanism.

Commonly, while studying electron transport through molecules, one assumes that
electrodes are large enough to have a feartureless electron density of states
below the Fermi energy. 
In the present work we analyze the NDR effect in the electron transport
through a molecular junction where the leads are small metal clusters whose 
electron density of states  reveals sharp and distinct features. This system 
provides better opportunities to analyze the effects of matching-mismatching 
of the molecule energy levels with those of the ``nanoleads" in the electron 
transport characteristics. We
show that under certain conditions such junctions may show very
distinguished multiple NDR features in the $I-V$ characteristics. This 
demonstrates their potential usefulness in nanoelectronics applications.

\begin{figure}[t]
\begin{center}
\includegraphics[width=1.7cm,height=6.5cm,angle=-90]{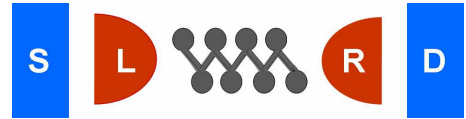}
\caption{ (Color online)
Schematic of the considered system including large electron reservoirs
labelled $S$  and $D,$ respectively; metal nano\-clusters $L $ and $R$;
and the molecule placed in between.
}
 \label{rateI}
\end{center}\end{figure}

To simplify the computational procedure we omit from the present consideration
effects originating from Coulomb interactions of electrons and from molecule 
vibrations. To maintain a steady supply of electrons tunneling through the
junction, we assume that the metal clusters (``nanoleads") keep contact with
large source and drain electron reservoirs as sketched in the
Fig. 1. The latter, however, are separated 
from the molecule is such a way that electrons cannot directly tunnel from 
these reservoirs to the molecule. In the absence of the applied bias voltage 
the whole system is supposed to be in the equilibrium state characterized with
the equilibrium Fermi energy $E_F.$ We write the effective Hamiltonian for the
molecule $(H_{eff})$ in the usual form \cite{1,19,20,21}:

  \be 
H_{eff} = H_M  + H_{L} + H_{R}  
     \label{1}
  \ee 
  where the term $H_M$ corresponds to the molecule itself, and $H_{L,R}$ 
 describe the coupling of the latter to the metal
nanoclusters.

Omitting for a while the molecule coupling 
to the leads, one may introduce the retarded Green's function for the 
molecule. The  latter is defined by the matrix 
equation:
     \be 
    \big[(E +i\eta) \hat S - \hat H_M \big] \hat G_0^R = \hat I
 \label{2}
  \ee
   where $I$ is the identity matrix, $ \hat S $ is the overlap matrix,
  \be 
  S_{ij} = \int \psi_i^* {\bf(r)} \psi_j {\bf(r)} d^3 r, 
  \label{3}      \ee
$ \eta $ is an infinitesimal positive parameter $(\eta \to 0^+)$, 
$ E $ is the energy,   and $ \psi_i, \psi_j $ are the orbitals included 
into the basis set.

Then we employ the Dyson equation. This equation relates the Green's function 
of the molecule coupled to the leads $\hat G^R (E) $ to the Green's 
function of the single molecule $ \hat G_0^R (E). $ It reads \cite{22}:
   \be 
  \hat G^R (E) =  \big\{ \big[\hat G^R_0 (E)\big]^{-1} - \hat\Sigma 
(E) \big\}^{-1}   . \label{4}
  \ee
  Here, the self-energy correction $\hat\Sigma(E)$ consists of two terms
describing the effect of two clusters:
  \be 
  \hat\Sigma (E) = \hat\Sigma_L (E) + \hat \Sigma_R (E).\label{5}
  \ee
 For convenience, one may introduce a notation $[\hat G_0^R(E)]^{-1} = \hat 
A(E),$ which allows to simplify the form of the Eq. \ref{4}, namely:
   \be 
 \hat G^R (E) = \big[\hat A(E) - \hat\Sigma(E)\big]^{-1}. \label{6}
  \ee
  Now, one must calculate matrix elements $A_{ij} (E) $ and $\Sigma_{ij}(E) $
 Assuming 
that the wavefunctions are orthonormalized molecular orbitals, 
the matrix $\hat A (E) $ is a diagonal matrix:
  \be 
   A_{ij} (E) = (E +i\eta - E_i) \delta_{ij} \label{7}
  \ee
  where $E_i $ are the energy eigenstates of the molecule. 

The matrix elements of self-energy corrections have the form \cite{1}:
   \be
 (\Sigma_\beta)_{i,j} = \sum_k
\frac{W_{ik,\beta}^*W_{kj,\beta}}{E-\epsilon_{k,\beta} - \sigma_{k,\beta}} .
   \label{8}  \ee
  here, $\beta \in L,R, \ W_{ik,\beta}$ are, respectively, the coupling 
strengths between $``i"$-th molecule state and $``k" $-th state on the 
left/right  metallic cluster (lead), $\epsilon_{k,\beta} $ are energy levels 
of the corresponding leads, and the parameters $\sigma_{k,\beta}$ are the
self-energy corrections which originate from the coupling of the clusters to
the large electron reservoirs. Their imaginary parts characterize the width
of the clusters energy levels. The summation over $``k" $ in the Eq. \ref{8}
is carried out over the states of the left/right cluster.

When the bias voltage $ V $ is applied across the system shown in the Fig. 1,
this causes charge redistribution, and subsequent changes in the energies $E_i$
and $\epsilon_{k,\beta}.$ In consequence, the matrix elements $ A_{ij} $ and 
$\Sigma_{ij} $ values vary as $V $ changes. This affects the electron transmission
function $T $ given by the expression:
   \be
T = Tr \big\{\hat\Gamma^L\hat G^R \hat\Gamma^R\hat G^A \big\} \label{9}
  \ee
  where $\hat\Gamma^{L,R} =- 2\mbox{Im}\hat\Sigma_{L,R},$ and $G^A $ is the 
advanced Green's function of the molecule $\big(\hat G^A = (G^R)^\dag\big)$. 
When the voltage $ V $ is applied, 
the electron transmission which determines transport properties of the
molecular junction may significantly depend on its value and polarity.

The electron tunnel current through the junction could be written in the form:
  \be
 I= \frac{e}{\pi\hbar}\int_{-\infty}^\infty dE T(E,V_{mol}) 
\left[f(E - \mu_S) - f(E - \mu_D)\right].     \label{10}
  \ee
 Here, $T(E,V_{mol}) $ is the electron transmission, given by the Eq. \ref{9},
$ f(E)$ is the Fermi distribution function for the energy $E.$ Chemical
potentials $\mu_{S,D}$ are attributed to the source and drain reservoirs,
respectively. They are shifted
with respect to the equilibrium Fermi energy due to the applied voltage:
 \be 
  \mu_S = E_F + (1- \nu) eV; \qquad \mu_D = E_F + \nu eV      
      \label{11}
  \ee
   where $ e $ is the electron charge and $ \nu $ is the division 
parameter. 

We remark that the transmission $ T $ in the Eq. \ref{10} actually depends
on the voltage applied across the junction $(V_{mol} )$. The latter may
noticeably differ from the external voltage $V.$ The difference originates
from both charge redistribution inside the molecule and the electrostatic
potential drops between the source/drain reservoirs and the metal clusters.
  The current-voltage characteristics shapes crucially depend on the 
electrostatic potential profile in the considered system. If the voltage
mostly drops between the large electron reservoirs and the small metal 
nanoparticles included in the junction $(V_{mol} \ll V)$ one may 
approximate $T(E,V)$ as $T(E,0)$. 
Then the applied bias 
voltage does not change relative positions of the energy levels 
$\epsilon_{k,\beta} $ and $E_i,$ and low-temperature 
characteristics should display step-like shapes. 
These are typical for electron tunneling through molecules (see e.g.
\cite{19,23}). Current increases as the voltage $ V $ increases, and the
NDR does not appear.
On the contrary, when no significant voltage drop occures between the
electron reservoirs and the metal nanoclusters $(V_{mol} \sim V),$ the
effect of the bias voltage on the energy levels of the clusters 
$\epsilon_{k,\beta}$ could considerably differ from its effect on the
molecular energy $ E_i.$ Due to these differencies in the voltage 
induced shifts, the relative positions of the energy levels on the
clusters and these on the molecule vary as the voltage changes. This 
creates opportunities for alignment/disalignment of the molecule orbitals
with those associated with the metal clusters. Therefore, in such a case 
one may expect the NDR to occur.

\begin{figure}[t]
\begin{center}
\includegraphics[width=4.3cm,height=9.4cm,angle=-90]{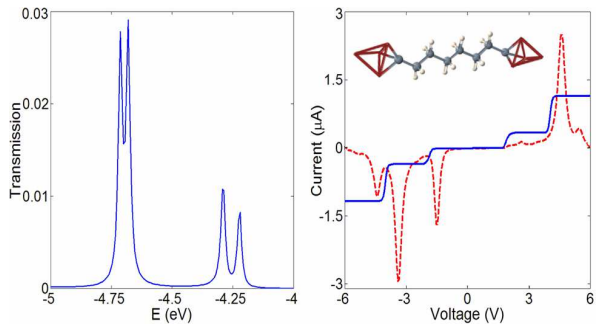}
\caption{ (Color online)  Left panel:
The calculated electron transmission function through the molecule $T(E,0)$
within a certain range of energies $E.$ Right panel:
The calculated current voltage chara\-cteristics for the electron tunneling
through the junction. The curves are plotted asuming $ T=50K, $ and
$ V_{mol} = 0.1 V $ (solid line) and $ V_{mol} = 0.9 V $  (dashed line).
}
 \label{rateI}
\end{center}\end{figure}

To confirm the above suppositions we carried out calculations of the tunnel 
electric current through a junction which consists of two copper
nanoclusters and an alaphatic-saturated hydrocarbon chain situated in 
between them (see Fig. 2). The relevant eigenenergies and matrix elements 
included in the Eqs. \ref{3}-\ref{9} are computed using the NRLMOL software 
package \cite{24}. 
First we performed full self-consistent
calculations on the considered nanosystem. 
In addition to solving the Kohn-Sham equations self
consistently we have optimized the geometry as well. Then we use Lowden's
method \cite{25} of symmetrical orthogonormalization to construct atom centered
wannier-like functions \cite{26} from the nonorthogonal atom-centered gaussian
orbitals. We reconstructed the Hamiltonian matrix in this basis
and block diagonalized it separating out the blocks coprresponding to the
hydrocarbon chain and the copper clusters.
Small off-diagonal elements between
the L-M and M-R blocks of the Hamiltonian describe the coupling of the molecule 
chain to the nanoleads.

The final matrix was used to determine the various matrices needed in
Eq. \ref{1}-\ref{11}.
In the following calculations of the tunnel current we 
assumed $ V_{mol} = 0.1 V$ and $ V_{mol} = 0.9 V,$ respectively. The 
resulting $ I-V $ curves are presented in the Fig. 2. One can see that when
the electrostatic potential mostly drops between the sourse/drain 
reservoirs and the nearby copper clusters $ (V_{mol} = 0.1 V),$ the 
corresponding  $ I-V $ curve reveals a stepwise profile without NDR features.
On the contrary, distinct NDR peaks emerge provided that there exists a
significant drop in the electrostatic potential between the copper clusters 
and the molecule tips $( V_{mol} = 0.9 V).$ Also, one may notice that some
NDR peaks are rather sharp and narrow, and the current peak values are much
greater than  in the valleys between the peaks. This is consistent
with the experimental data reported in the  \cite{8}.

In conclusion, we have considered a metal-molecule-metal
tunneling junction where metal leads are nanoparticles. 
Due to the extremely small 
size of these metal clusters their electron density of states reveals distinct
features which are washed out for larger electrodes. We did show that 
 NDR peaks could appear due to alignment and subsequent disalignment
of the energy levels of the metal nanoclusters with those of the molecule
when the voltage applied across the junction varies. The proposed analysis
is based on simple assumptions concerning the electrostatic potential
distribution inside the junction. It seems unlikely 
that these simplifications could cause qualitative distortions
in the considered NDR manifistations. However, to quantitatively analyze the
effect one must properly compute the electrostatic potential profile 
employing a self-consistent computational procedure. Also, one should take 
into account the effects of molecular vibrations which could result in extra 
NDR features superimposed upon those presently analyzed. Nevertheless, we do
believe that we showed that the tunneling junctions including a molecule
placed in between metallic nanoparticles could exhibit distinct NDR peaks
in the $ I-V$ curves. This makes such junctions  useful 
is designing nanoelectronic devices.

{\it Acknowledgments:} We thank  G. M. Zimbovsky for help 
with the manuscript. NZ acknowledges support from the ASEE and ONR Summer Faculty Research Program.


\begin{thebibliography}{99}

\bibitem{1} S. Datta, {\it Quantum Transport: Atom to Transistor}
 (Cambridge University Press, 2006).


\bibitem{2} See e.g. G. Cuniberti et al (Ed), {\it Introduction to Molecular
Electronics} (Springer, Berlin, 2005).

\bibitem{3} L. Ezaki and P. J. Stiles, Phys. Rev. Lett.  {\bf 6}, 1108 (1966).

\bibitem{4} L. L. Chang, E. E. Mendez, and C. Tejedor, {\it Resonant Tunneling
in Semiconductors} (Plenum, New York, 1991). 

\bibitem{5}M. A. Kastner, Rev. Mod. Phys. {\bf 64}, 849 (1992).

\bibitem{6}  L. W. Yu, K. J. Chen, J. Song. J. M. Wang, J. Xu. W. Li, and X. F. 
Huang, Thin Solid Films {\bf 515}, 5466 (2007).

\bibitem{7} J. Gandioso, L. J. Lauhon, and W. Ho. Phys. Rev. Lett. {\bf 85}, 1918,
(2000).

\bibitem{8} F.-R. F. Fan, R. Y. Lai, J. Cornil, Y. Karzari, J-L. Bredas, L.-T. 
Cai, L. Cheng, Y. Yao, D. W. Price, Jr., S. M. Dirk, J. M. Tour, and A. J. Bard,
J. Am. Chem. Soc. {\bf 126}, 2568 (2004).

\bibitem{9} A. Salomon, R. Arad-Yellin, A. Shanzer, A. Karton, and D. Cahen,
J. Am. Chem. Soc. {\bf 126}, 11648 (2004).

\bibitem{10} M. Grobis, A. Wachowiak, R. Yamachica, and M. F. Crommie, Appl.
Phys. Lett. {\bf 86}, 204102 (2005).

\bibitem{11} J. Repp, G. Meyer, S. M. Stojkovic, A. Gourdon, and C. Joachim,
Phys. Rev. Lett. {\bf 94}, 026803 (2005).

\bibitem{12} J. J. Davis, T. Wang, A. Morgan, G. Zhang, J. Zhao, Faraday
Discuss. {\bf 131}, 167 (2006).
 
\bibitem{13} E. D. Mentovich, I. Kalifa, A. Tsukernik, A. Caster, N. 
Rosenberg-Shraga, H. Marom, M. Gozin, S. Richter, Small {\bf 4}, 55 (2007).

\bibitem{14}  G. Maruccio, P. Marzo, R. Krahne, A. Passaseo, R. Cingolani, and
R. Rinaldi, Small {\bf 3}, 1184 (2007).



\bibitem{15} Y. Xue, S. Datta, S. Hong, R. Reifenberger, J. I. Henderson, and
C. P. Kubiak, Phys. Rev. B {\bf 59}, R7852 (1999).

\bibitem{16} N. Simonian, J. Li, and K. Likharev, Nanotechnology {\bf 18}, 424006
(2007).

\bibitem{17} V. Mujica, A. Nitzan, S. Datta, M. A. Ratner, and C. P. Kubiak,
J. Phys. Chem. B {\bf 107}, 91 (2003).



\bibitem{18} M. Yu. Galperin, M. A. Ratner, and A. Nitzan, Nanoletters {\bf 5},
125  (2005).

\bibitem{19} N. A. Zimbovskaya and G. Gumbs, Appl. Phys. Lett. {\bf 81}
1518 (2002).

\bibitem{20} Y. Xue, S. Datta, and M. A. Ratner, J. Chem. Phys. 
{\bf 115}, 4292 (2001).

\bibitem{21} M. Yu. Galperin, M. A. Ratner, and A. Nitzan, J. Chem. Phys.
{\bf 121}, 11965 (2004).

\bibitem{22} M. Yu. Galperin, A. Nitzan, and M. A. Ratner, Phys. Rev. Lett.
{\bf 86}, 166803 (2006). 


\bibitem{23} V. Mujica, A. E. Roitberg, and M. A. Ratner, J. Chem. Phys. 
{\bf 112}, 6834 (2000).

\bibitem{24} M. R. Pederson, D. V. Porezag, J. Kortis, and D. C. Patton,
Phys. Status Solidi B {\bf 217}, 197 (2000).


\bibitem{25} P. O. Lowden, J. Chem Phys. {\bf 18}, 365 (1950).


\bibitem{26} M. R. Pederson and C. C. Lin, Phys. Rev. B {\bf 35}, 2273 (1985).


\end{thebibliography}
\end{document}